# Accurate strain measurements in highly strained Ge microbridges

A. Gassenq,[1,2], S. Tardif,[1,2] K.Guilloy,[1,2] G. Osvaldo Dias,[1,3] N. Pauc,[1,2] I. Duchemin,[1,2] D. Rouchon,[1,3] J-M. Hartmann,[1,3] J. Widiez,[1,3] J. Escalante,[1,2] Y-M. Niquet,[1,2] R. Geiger,[4] T. Zabel,[4] H. Sigg,[4] J. Faist,[5] A. Chelnokov,[1,3] F. Rieutord,[1,2] V. Reboud,[1,3] V. Calvo[1,2]

[1]*Univ. Grenoble Alpes, 38000, Grenoble, France*

[2]*CEA-INAC, 17 rue des Martyrs, 38000, Grenoble, France*

[3]*CEA-LETI, Minatec Campus, 17 rue des Martyrs, 38054, Grenoble, France*

[4]*Laboratory for Micro- and Nanotechnology, Paul Scherrer Institut, 5232, Villigen, Switzerland*

[5]*Institute for Quantum Electronics, ETH Zurich, 8093, Zürich, Switzerland*

Ge under high strain is predicted to become a direct bandgap semiconductor. Very large deformations can be introduced using microbridge devices. However, at the microscale, strain values are commonly deduced from Raman spectroscopy using empirical linear models only established up to $\varepsilon_{100}$=1.2% for uniaxial stress. In this work, we calibrate the Raman-strain relation at higher strain using synchrotron based microdiffraction. The Ge microbridges show unprecedented high tensile strain up to 4.9 % corresponding to an unexpected $\Delta\omega$=9.9 cm$^{-1}$ Raman shift. We demonstrate experimentally and theoretically that the Raman strain relation is not linear and we provide a more accurate expression.

## 1. INTRODUCTION

Strain engineering has become a widely used strategy to enhance the performance of semiconductor devices[1,2] such as transistors[3,4], modulators[5], piezoelectrics[6] and semiconductor lasers[7], to name a few. Many material properties like electronic band structures can indeed be tailored by strain. For Germanium, it has been predicted that high tensile strain can tune the relative band gap energies, improving light emission and transforming it into a direct band gap material,[8–12] opening the way to mid-infrared lasers fully compatible with Complementary Metal Oxide Semiconductor (CMOS) technology. The tensile strain needed to reach a direct bandgap has been theoretically estimated to be around 4.6 %[13–15] for a uniaxial loading along <100>. Reaching such large strain values while retaining crystal integrity is extremely challenging. Several methods are currently being explored[16–19] and the highest strains are obtained by strain redistribution[13,20]. The attained strains exceed in a radical way the intrinsic strain limits of conventional epitaxial growth[21,22].

The method of choice for a direct, model-free determination of strain in a crystalline material is X-ray diffraction. At synchrotrons, latest developments allow strain measurements down to the sub-micrometer scale[23–25]. However, due to the accessibility, micro-Raman spectroscopy is routinely used in the laboratory to quantify the local strain from a measurement of the peak frequency shift of an optical phonon mode. To determine the strain, experimental strain-to-frequency conversion rules are used. The rules are specific for each material and orientation of the strain.[26,27] For Germanium (Ge), the Raman shift conversion rules were experimentally established for uniaxial strain up to 1.2 %[27] which is much lower than the theoretical ideal yield strength of monocrystalline Ge of approximately 18 %.[28,29] In practice, crystalline defects and roughness at the interfaces reduce the limit of rupture. Nevertheless, the strains achieved now in Ge micro-structure are several-fold higher[13,20] than what was previously studied to establish the Raman-strain relation. Therefore, the Raman strain shift coefficient needs to be calibrated at higher level of strain in order to give access to accurate high strain micro-measurements in Ge with laboratory equipment.

In this work, we demonstrate accurate measurements of elevated strain in Ge via synchrotron based X-ray micro-diffraction and calibrate the strain-to-Raman-shift up to tensile strain of an unprecedented level of 4.9 %, corresponding to 9.9 cm$^{-1}$ Raman wavenumber shift.

## 2. Strain induction using microbridge devices

Stress in the semiconductor layer is applied in a controllable way by strain redistribution. The microbridge geometry was initially proposed in Si[30] and later in Ge[19]. At that time[19], only limited strain (Raman shifts up to 4.8 cm$^{-1}$) was reached in Ge before fracturing the microbridges, most likely due to the presence of an array of misfit dislocations at the Ge/Si interface. For this work, we used high crystalline quality 200-mm optical Ge-on-Insulator (GeOI) wafers fabricated by Smart-Cut™ technology[31–34] in order to shift the mechanical failure of microbridges to higher strain.[20,31] Fig. 1(a) presents the process flow used to fabricate Ge microbridges (Fig. 1(b)). Ge layers are 0.35 µm thick with a biaxial strain evaluated at 0.16 %. Microbridge patterning was performed using e-beam lithography and Ge etching in an inductively-coupled plasma reactor. The Ge membranes were then released using under-etching in a dedicated etching reactor combining anhydrous HF vapors and alcohol vapors. Thanks to the under-etching, tensile strain was concentrated in the narrowest part of the suspended microbridge. The strain homogeneity in the central region allows measuring locally the material with 1 µm-diameter probes.[35]

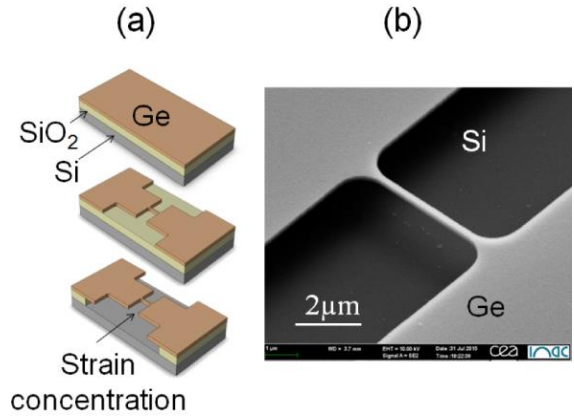

FIG. 1: (a) Process flow used for the fabrication of suspended microbridges from optical GeOI wafers; (b) Fabricated Ge microbridges

## 3. Strain measurement

The strain state in the Ge microbridges was measured using X-ray Laue and rainbow-filtered Laue micro-diffraction at beam-line BM32 of the European Synchrotron Radiation Facility in Grenoble.[35,36] The energy spectrum of the white synchrotron X-ray beam ranged from 5 to 25 keV and the beam was focused (with a 0.5 µm x 0.5 µm spot size) on the central part of the microbridges. Laue diffraction patterns were collected on a 2 dimensional MARCCD165 detector. For rainbow-filtered measurements, a diffracting diamond plate was inserted in the beam path to control the incident energy spectrum.

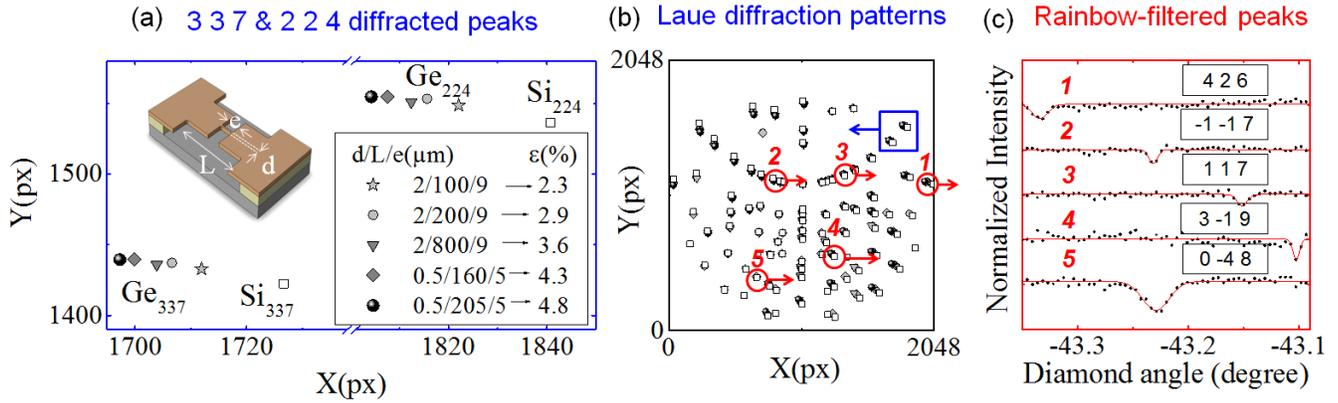

FIG. 2: (a) Zoom on the 3 3 7 and 2 2 4 Bragg reflections from the (b) full Laue diffraction pattern of the superimposed diffraction patterns from the Si substrate and several microbridges with different dimensions; (c) Intensity of 5 selected Bragg reflections as a function of the diamond plate angle in the rainbow-filtered µLaue technique for the membrane with the 2.9 % strain.

Typical Laue diffraction patterns for different microbridges on the same chip are shown in Fig. 2(a) and (b). Fig. 2(a) is a zoom of Fig. 2(b) showing the evolution of two diffraction spots corresponding to the Miller indices of the 337 and 224 Bragg reflections of the Si substrate and of several suspended Ge microbridges processed on the same chip. Depending on the bridge design (parameters d, L and e indicated in the inset), a shift of the Ge spot positions is observed. The full widths at half maximum values are constant (1.3 ± 0.1pixels) which indicate that the strain stays homogeneous when the deformation increases. The deviatoric strain tensor was calculated from the peak positions on the Laue pattern using the software LaueTools.[37] The longitudinal strain value was obtained from the deviatoric strain tensor by considering no stress normal to the free surfaces.[35] Measured strains are indicated in the legend; 4.8 % is achieved in this device. A maximum of 4.9 % will be presented hereafter in even narrower microbridges (250 nm width). Additional direct measurements of the longitudinal strain were performed on a few microbridges using the rainbow filter[36]. Only few microbridges carefully chosen over the whole strain range were measured during the limited time span of the beam-time. The rainbow-filtered measurement in a 2.9 % strained membrane is shown as an example in Fig. 2(d), where the intensities of 5 different Bragg reflections are plotted as a function of the angle of the diamond plate. The corresponding Bragg reflections on the detector are indicated by circles in Fig. 2(b). The intensity drops when the energy of the Bragg reflection corresponds to a diffraction of the diamond plate which allows to measure directly the lattice parameter. Knowing the energy of the Bragg reflection provides the value of the spacing of the atomic planes.[36] We were thus able to locally access the strain in Ge microbridges using two micro-diffraction techniques.

## 4. Raman spectroscopy

As far as strain micro-characterization is concerned, Raman spectroscopy has the advantage of being relatively well spatially resolved, fast and widely available in laboratories. Under optical excitation, lattice vibrations generate a variation in the electrical susceptibility of the Ge crystal, which gives rise to Rayleigh and Raman scattering. Mechanical stress affects the spectral frequencies of the Raman modes, giving access to the induced strain. The relationship between strain and Raman wavenumber shift will be detailed hereafter. In this work, a micro-Raman spectrometer with a 532 or 785 nm wavelength excitation laser is used to probe the strain in Ge microbridges. The light was focused on the sample surface with a 100x short working distance objective. Measured Raman spectra were compared for both wavelengths. Since the measured Raman spectral shifts were similar for low input power, the 785 nm wavelength was chosen for a deeper probing depth (~200 nm [38]) with a resulting spot diameter around 1 μm. Power dependence measurements were performed in order to quantify and correct heating effects.[39] From this calibration, the laser intensity, fixed at 9 μW, was focused on the sample. The heating effect was lower than the indicated 0.1 cm$^{-1}$ uncertainty coming from the Raman spectrometer. The Raman spectral shift was measured by fitting the Raman spectra with Lorentzian functions. A bulk Ge (001) substrate was systematically used as a reference for 0 % strain. Fig. 3 presents the measured spectra for several micro-bridges. The typical dimensions (L, d and e) used to tune the strain are indicated in the scale. For d = 250 nm, the spectral shift reaches 9.9 ± 0.1 cm$^{-1}$, which is the highest reported value in the literature up to date for Ge micro-bridges.

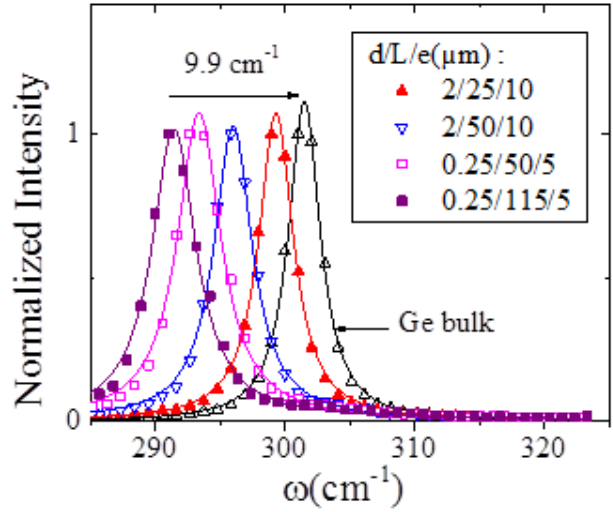

FIG. 3.  Raman spectra of different Ge microbridges

**5. Raman-Strain relation**

We can now link such Raman spectral shifts to the strain obtained from X-ray micro-diffraction measurements performed on the same microbridges. Fig. 4 presents the measured strain as function of the spectral Raman shift. The strains exhibits a pure uniaxial stress confirmed by a measured strain ratio between the different crystallographic orientations defined by the elastic coeficient.[40] Diffraction (circles) and rainbow-filtered Laue diffraction (stars) measurements exhibit a very good agreement which confirms the absence of stress normal to the free surfaces. A clearly nonlinear relationship is found, with the maximal strain evaluated at 4.9 ± 0.1 % for a shift of 9.9 ± 0.1 cm$^{-1}$.

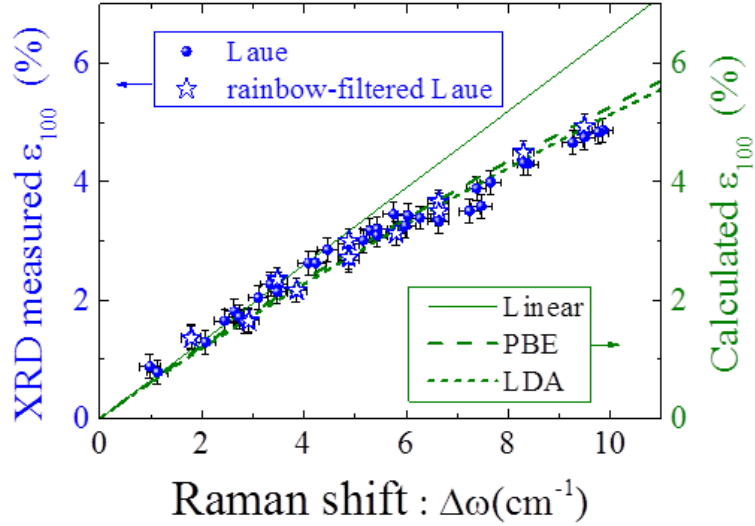

Fig. 4. Raman strain relation: measured strain by micro-XRD (blue circles and stars), simulated strain by ab initio calculations (dotted green lines), and linear empirical dependence (solid green line)

Till now, linear relationships were widely used in the literature to link Raman shift to strain in Ge.[27,41] The linearity was experimentally confirmed for several situations with relatively low strain (up to 1.2 % uniaxial strain in Ge[27]; up to 2.6 % bi-axial strain in Ge [42–44]). For uniaxial stress along <100> in Ge, the commonly used linear empirical relation is given by Equation 1 with ε in % and ω in $cm^{-1}$.[19,27] The standard error on the strain-shift coefficient can be evaluated to approx. 0.02 coming from the phonon deformation potential uncertainty.[27]

$$\Delta\varepsilon_{100} = 0.65(\pm 0.02) \times \Delta\omega \quad (1)$$

The deviation from linearity of the Raman shift versus strain for the <100> loading is evident from the data shown in Fig. 4. To provide the revisited strain-shift relation over the 0-5 % tensile strain and 0-10 $cm^{-1}$ shift range, we have fitted our experimental values by the Equation 2. The indicated uncertainties are evaluated by the fit deviation.

$$\Delta\varepsilon_{100} = 0.68(\pm 0.02) \times \Delta\omega - 0.019(\pm 0.002) \times \Delta\omega^2 \quad (2)$$

Taking into account the coefficient uncertainties, we find a good agreement between Eq.1 and Eq.2 at low strain values (<2.5%), which is expected. At higher strains the difference between Eq. 1 and Eq. 2 is substantial: for example, from Eq. 1 we obtain that a Raman shift of 9.9 $cm^{-1}$ would be converted to 6.4 % strain, while the actual strain value is 4.9 %, instead. To support the experimental findings, we performed ab-initio calculations of the Raman shifts. The dependence of the TO phonon wavenumber on strain, was determined up to 5.5 % from both Local-Density Approximation (LDA) and Perdew-Burke-Ernzerhof (PBE) exchange correlation functionals using the Abinit Density Functional Theory (DFT) package[45–47]. In order to circumvent metallicity problems arising from DFT calculation in crystalline Ge, we enforced a complete occupation of the valence band structure over the whole Brillouin zone, corresponding effectively to a 0K electronic temperature. The relaxation of the strained crystalline structure along the axis perpendicular to the constraint has been obtained for the 8 atoms orthorhombic supercell. Response function phonon calculations were performed on the resulting relaxed 2 atoms unit cell. For TO phonons which are measured in our experiments, PBE and LDA calculations give similar results and exhibit a very good agreement to the experimental values (Fig. 4) which supports the new provided Raman-strain relation (Eq.2). As a consequence, reported high strain values in the pioneer works[13,15,19] using the linear model (Eq.1) for high levels of strain are overestimated.

## 6. Conclusion

To sum up, thanks to the high crystalline quality of 200 mm optical GeOI wafers, large uniaxial strain in suspended Ge microbridges was achieved up to a record-breaking 4.9 %, corresponding to a Raman spectral shift of 9.9 cm$^{-1}$. A large range of strain was measured by X-ray micro-diffraction and the relationship between experimental Raman shifts and strain was determined. The obtained unexpected deviation from the linear dependence was confirmed by ab-initio calculations. A new relationship between Raman shift and strain is provided. It allows for accurate measurements of extreme strain in Ge with Raman spectroscopy, which is a key for the realization of a direct bandgap material for CMOS compatible laser application.
**Acknowledgements**

**Acknowledgements**

The authors would like to thank the Platforme de Technologie Amont in Grenoble for clean room facilities and the beamline BM32 at ESRF for synchrotron based measurement. This work was supported by the CEA DRF-DRT Phare projects "Photonics" and "Operando", the CEA-Enhanced Eurotalent project "Straintronics" as well as the Swiss National Science foundation SNF.